\documentstyle[twocolumn,eqsecnum,aps]{revtex}
\include{epsf}

\begin{document}
\draft
\preprint{}
\title{Electrostatic force spectroscopy of a single InAs quantum dot
}
\author{Aykutlu D\^{a}na, Charles Santori and Yoshihisa Yamamoto}
\address{Quantum Entanglement Project, ICORP, JST\\
Edward L. Ginzton Laboratory, Stanford University, Stanford, California 94305-4085}
\date{\today}
\maketitle

\begin{abstract}
Electrostatic force microscopy at cryogenic temperatures was used
to probe the electrostatic interaction between a conductive atomic
force microscopy tip and electronic charges trapped in an InAs
quantum dot. Measurement of the self-oscillation frequency shift
of the cantilever as a function of tip-sample bias voltage
revealed discrete jumps at certain voltages when the tip was
positioned above a quantum dot. These jumps are attributed to
single-electron filling of the electronic states of the quantum
dot. The estimated resonant energies of two s-states and four
p-states are compared with the experimental data obtained for an
ensemble of quantum dots by the capacitance measurement technique.
\end{abstract}

\section{introduction}

Self assembled InAs quantum dots (QDs) grown on a GaAs substrate
have attracted much attention recently due to their high crystal
quality and numerous potential applications. Optical and
electrical spectroscopy techniques to measure the discrete energy
levels of InAs QDs have  been
developed\cite{marzin,bayer,ribeiro,ribeiro2}. Optical spectroscopy allows
high energy resolution, and individual QDs can be studied by
spatially isolating them. This technique probes only the allowed
transition of electron-hole pair. Capacitance spectroscopy, on the
other hand, can probe electron and hole resonant energies
seperately, but has so far been performed only on structures
containing large numbers of quantum dots, due to the lack of
spatial resolution and sensitivity. Recently, scanning capacitance
microscopy in contact mode has been applied to capacitance imaging
of QDs without evidence for single charge effects \cite{scmqds}.

Application of atomic force microscopy (AFM) to measurement of electrostatic
force gradients due to few electronic charges trapped below the surface
has been demonstrated \cite{berkowitz1,mcgill}. In these experiments,
only indirect estimations of the electron energy levels were given based on
tunneling rate from the localized states into the substrate. Measurement
of frequency shift of a self-oscillating cantilever
as a function of bias voltage has also been used to look at sub-surface charge
densities without indication of single charge effects \cite{guggisberg1}.

In this letter, we report, a new measurement technique to detect
trapped charge in individual InAs QDs. A non-contact measurement
scheme is employed, where a conductive
AFM tip is positioned above a QD to probe the gradient of the
electrostatic force between the tip and charges trapped in the QD
as a function of tip-sample potential difference.

\section{Experiment}

The sample used in the experiment was grown on an n-type GaAs
(100) substrate by MBE. A 0.5 $\rm{\mu}m$ thick GaAs buffer layer
with $10^{18}$ $\rm{cm^{-3}}$ Si doping was first grown, followed
by the growth of a semi-insulating GaAs tunnel barrier of 15 nm
thickness. After growth of a monolayer InAs wetting layer, InAs
QDs with average base diameter of 20 nm were grown on top of the
tunnel barrier. The density of the QDs was observed to be about
${10^{10}}$ ${\rm cm^{-2}}$ at the center of the wafer and
decreasing towards the edges. The homemade AFM used in this
experiment used a fiber-interferometric deflection sensor for
cantilever readout, with a wavelength of 1310 nm and an incident
light power of about 5 $\rm{\mu}$W.

\begin{figure}
\begin{center}
\epsfxsize=2in 
\epsffile{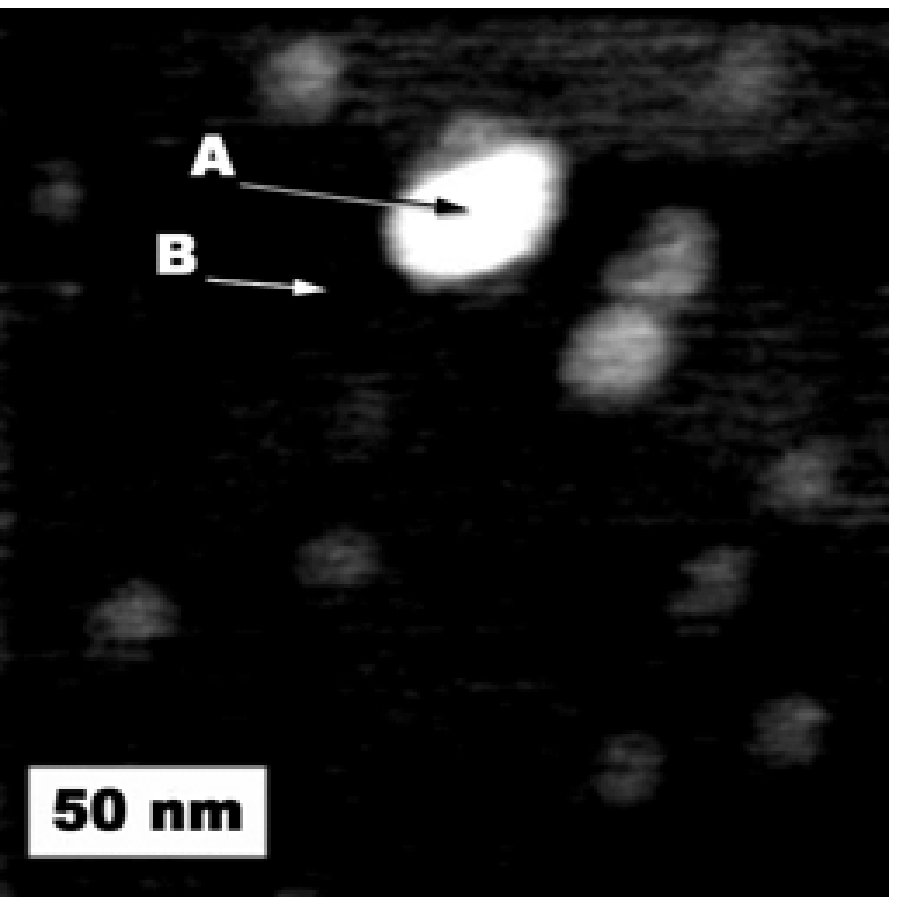}

\epsfxsize=2.75in 
\epsffile{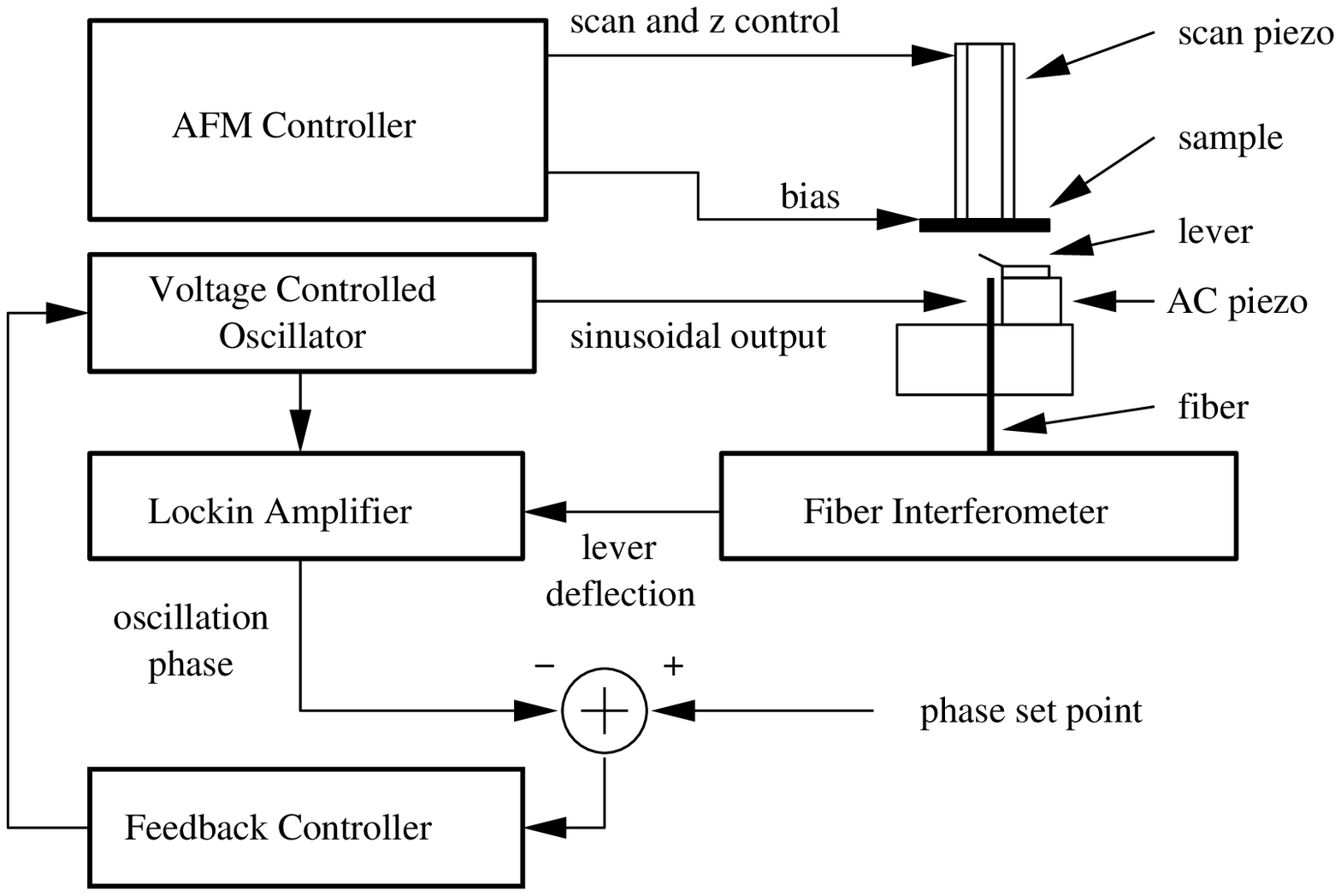}

\end{center}
\caption{(a) Contact mode image of InAs quantum dots. Locations
where electrostatic force spectroscopy is performed are labeled A
and B. (b) Schematic of the experimental setup.}\label{fig1}
\end{figure}

The Pt/Ir coated AFM cantilever had a spring constant of 2.5 N/m
and a fundamental resonance frequency of 68.8 KHz. The Q factor of
the cantilever was about 28,000 at 77K and increased to 59,000 at
4.2 K.  Contact mode images of QDs were obtained prior to
electrostatic force spectroscopy (EFS) on individual dots. Imaging
and EFS were done at a temperature of 4.2 K. Since there was no
capping layer on the QDs, base diameters and heights can be
accurately determined from the contact mode images (Fig. 1a).

After contact mode imaging, the tip was lifted 10 to 30 nm above
the surface. Self oscillation of the cantilever was sustained
through an analog tracking oscillator circuit \cite{guentherodt}.
The circuit is a phase-locked loop (PLL) system that drives the
cantilever at its resonant frequency with a fixed drive amplitude.
Tracking of the frequency is accomplished by locking the local
oscillator phase to the AFM cantilever oscillation phase by a
proportional-integral feedback controller. The block diagram of
the EFS setup is shown in Fig. 1b. During the EFS measurements,
the oscillation amplitude was fixed to be about 1.5 nm, to provide
adequate signal-to-noise ratio while maintaining a small ratio
between the oscillation amplitude and the tip-sample seperation.
In this regime, the linear approximation used in the theoretical
calculation of the frequency shifts is justified. The minimum
detectable frequency shift is fundamentally limited by the
thermomechanical fluctuation of the cantilever and is given by
\begin{eqnarray}
{\rm{\delta f_{min}
=\sqrt{{f_0k_BTB}\over{2\pi Qk{A_{osc}^2}}}}}
\end{eqnarray}

where $\rm{f_0}$ is the cantilever resonance frequency,
$\rm{k_BT}$ is the thermal energy, Q is the mechanical quality
factor, k is the spring constant, $\rm{A_{osc}}$ is the amplitude
of cantilever oscillation and B is the measurement bandwidth. The
detectable frequency shift was also limited by the noise sources
in the secondary detectors such as optical interferometer and
electronic amplifier noise, and the overall sensitivity of our
system was about 0.5 Hz/$\rm{\sqrt{Hz}}$.

Since the tip radius is about 20 nm and the EFS is performed at a
tip sample separation of about 15 nm on average, a parallel plate
capacitor model can be used to analyze the experimental data. The
force between the tip and sample in the presence of a trapped
charge Q in the QD is given by (Refs. \cite{berkowitz1,mcgill}):

\begin{eqnarray}\nonumber
{\rm{F_{e}(z)={1 \over {(z+(d_1+d_2)/\epsilon_r)^2} }}}\\
 \times{
\left(-{{{d_1}^2Q^2} \over {\epsilon_r^2\epsilon_0A}}+
{{2{d_1}Q{V_{\rm {ts}}}} \over
{\epsilon_r}}+{{\epsilon_0A{V_{ts}}^2} \over {2}}\right).}
\label{electrostaticforce}\end{eqnarray}

Here, A is the effective tip area, ${\rm {d_1}}$ is the tunnel
barrier thickness, $\rm d_2$ is the optional capping layer
thickness, $ \rm {V_{\rm{ts}}}$ is the tip-sample bias voltage, ${\rm
{\epsilon_r}}$ is the relative dielectric constant of a GaAs
tunnel barrier, and z is the tip to sample-surface separation. A
WKB calculation based on a bulk and QD band diagram suggests that
the tunneling rate into the QD is fast enough that the QD is
charged and discharged at a time scale much faster than other time
scales involved in the experiment. Therefore, we assume that as
the tip-sample voltage is swept, the dot and the n-type reservoir
are in equilibrium with each other.

Inspecting individual terms of the electrostatic force, Eq.
(\ref{electrostaticforce}), it is seen that only the first and
second terms of the sum contains information about the localized
state and the third term merely provides a background
electrostatic force from the substrate. For experimental
conditions of interest, it is also seen that the second term of
the sum is two orders of magnitude larger than the first term,
therefore the first term can be neglected. Including only the
dominant term, we can approximate Eq. (\ref{electrostaticforce})
for a number of trap states that are at distances $\rm
h_{\textit{i}}$ above the substrate with individual charges $\rm
q_{\textit{i}}$ as

\begin{eqnarray}
{\rm{F_{e}(z) ={1 \over {(z+d/\epsilon_r)^2}} }
\sum_{\textit{i}}{{2{h_{\textit{i}}{q_{\textit{i}}}{V_{\rm {ts}}}}
\over {\epsilon_r}}}} \label{individualforces}\end{eqnarray}

where d is the overall dielectric thickness covering the
$\rm n^+$ substrate. To calculate the trapped charge $\rm q_{\textit{i}}$
for a given state \textit{i} as a function of tip-sample bias
voltage, it is assumed that the quantum dot is in equilibrium with
the substrate electron distribution. For state \textit{i}, the
charge in the state can be calculated by the Fermi-Dirac
distribution

\begin{eqnarray}
{\rm{ q_{\textit{i}} = e[1+\exp(
{E_{\textit{i}}-E_{f}\over{k_BT}})]^{-1}}}
\label{charge_i}\end{eqnarray}

In the parallel plate approximation, ignoring band bending effects
at the interfaces, energy level ${\rm {E_{\textit{i}}}}$ can be expressed
with reference to the bulk fermi level as the sum of a bias
induced energy shift and an inherent energy $\rm{E_{\textit{i},0}}$.
Assuming the bulk fermi level $\rm E_f$ to be the conduction band
energy $\rm E_{CB}$ for the $\rm n^+$ substrate, this bias dependent
energy for state \textit{i} is given by

\begin{eqnarray}
{\rm {E_{\textit{i}}(V_{ts})=E_{\textit{i},0}+V_{ts}
\frac{eh_{\textit{i}}}{\epsilon_r z+d}+E_{CB}.}}
\label{biasdependentenergy}\end{eqnarray}

As can be seen in Eq. (\ref{biasdependentenergy}), $\rm
E_{\textit{i},0}$ is the energy of state \textit{i} with respect
to the bulk conduction band $\rm E_{CB}$ under the zero bias
condition. The voltage dividing ratio which multiplies $\rm
V_{ts}$ in Eq. (\ref{biasdependentenergy}) can be deduced from the
thicknesses of the barrier layers and tip-sample separation. In
our experiments this ratio was typically 0.03 to 0.1.

During the course of EFS, the resonance frequency of the
cantilever is shifted in the presence of the external
electrostatic force gradient. This shift ${\rm {\delta}f}$ can be
calculated in the limit of small oscillation amplitude as

\begin{eqnarray}{\rm{\delta f(z)
={{f_0} \over {k_0A_{osc}}} \int\limits_{0}^{\,\,\,\,\,\,\,2\pi}
{{d\phi}\over{2\pi}}F(z+A_{osc}\rm cos \phi)\rm cos
\phi}}.\end{eqnarray}

One can approximate this frequency shift ${\rm {\delta}f}$ as

\begin{eqnarray}{\rm{\delta f={{f_0} \over {2k_0}} \langle
\frac{dF_e(z)}{dz}\rangle|_{A_{osc}}}}\end{eqnarray}

where the bracket denotes the average of the electrostatic force
gradient over the oscillation amplitude of the cantilever.

In order to isolate the discrete charging signature, we look at
the derivative of the frequency shift with respect to the tip
sample voltage
\begin{center}
\begin{eqnarray}

\nonumber \rm{\frac{\partial}{\partial V_{ts}}\delta f
=\frac{f_0}{2k_0A_{osc}}}\\
 \times{\frac{\partial}{\partial
V_{ts}}[F_e(z+A_{osc}/2)-F_e(z-A_{osc}/2)]}

\label{equationddfdv}
\end{eqnarray}
\end{center}

To calculate the contribution of electron state \textit{i} to the
derivative in Eq. (\ref{equationddfdv}), inserting
$\rm{F_{e,\textit{i}}(z)}$ from Eq. (\ref{individualforces}) and
neglecting small terms, we get

\begin{eqnarray}\nonumber
\rm {\frac{\partial}{\partial V_{ts}}\delta
f_{\textit{i}}}\simeq\frac{h_{\textit{i}}f_0V_{ts}}
{A_{osc}k_0\epsilon_r(z+d/{\epsilon_r})^2}\\
 \times{
\frac{\partial}{\partial
V_{ts}}[q_{\textit{i}}(z+A_{osc}/2)-q_{\textit{i}}(z-A_{osc}/2)]}
\label{peaks}\end{eqnarray}

In the zero temperature limit, Fermi-Dirac distribution which
determines the charging of the electron states approaches a step
function. Therefore, in this limit,  Eq. (\ref{peaks}) predicts
two delta functions for each state \textit{i} below the tip,
appearing at tip-sample voltages that depend on tip-sample
separation, overall semi-insulating layer thickness, the
oscillation amplitude of the cantilever as well as the spatial
location of the state and its energy with respect to the bulk
fermi level of the $\rm n^+$ GaAs region. The peaks in
$\rm{\partial(\delta f)/{\partial V_{ts}}}$ occur when the
condition

\begin{eqnarray}{\rm{E_{\textit{i},0}=-
\frac{eV_{ts}h_{\textit{i}}}{\epsilon_r(z\pm A_{osc}/2)+d}}}
\label{peakcondition}\end{eqnarray}

is satisfied.

In the finite temperature limit, further information can be
obtained from Eq. (\ref{peaks}) as it estimates two peaks of
opposite sign with magnitudes and widths determined by the height
$\rm{h_{\textit{i}}}$ of level \textit{i} above the $\rm {n^+}$
substrate. The derivative $\rm{\partial
q_{\textit{i}}(z,V_{ts})/{\partial V_{ts}}}$ is

\begin{eqnarray}\nonumber
{\rm {\frac{\partial}{\partial V_{ts}}q_{\textit{i}}(z,V_{ts}) =
\frac{e^2 h_{\textit{i}}}{kT(\epsilon_rz+d)}}}\\
 \times{
\frac{\exp[(E_{\textit{i}}(V_{ts})-E_{f})/kT]}{\{1+\exp[(E_{\textit{i}}(V_{ts})-E_{f})/kT]\}^2}}\end{eqnarray}

The width of the individual peaks can be approximated by

\begin{eqnarray}{\rm{\Delta V_{ts}=
\frac{2kT(\epsilon_r z + d)}{eh_{\textit{i}}}}}
\label{peakwidths}\end{eqnarray}

By analyzing EFS data for different tip-sample surface separations
and cantilever oscillation amplitudes in a parameter range where
the parallel plate approximation is valid, using Eq.
(\ref{peakcondition}) and Eq. (\ref{peakwidths}) one can solve for
$\rm{h_{\textit{i}}}$ and $\rm{E_{\textit{i}}}$ and extract
information on the location and the energy of states. Repeating
the procedure at multiple locations on the sample, a 3D mapping of
the localized density of states can be obtained.

\begin{figure}
\begin{center}
\epsfxsize=3in 
\epsffile{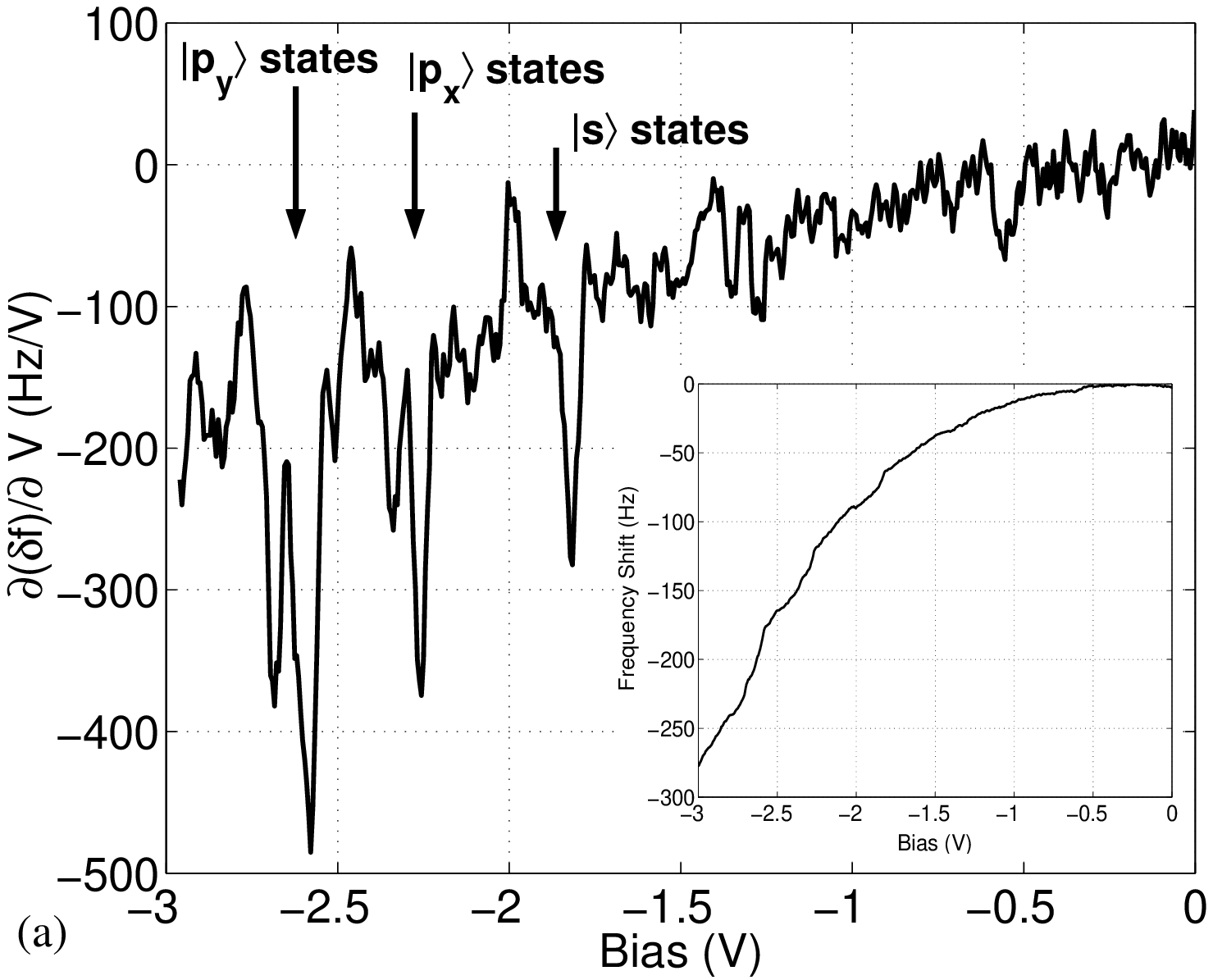}
\epsfxsize=3in 
\epsffile{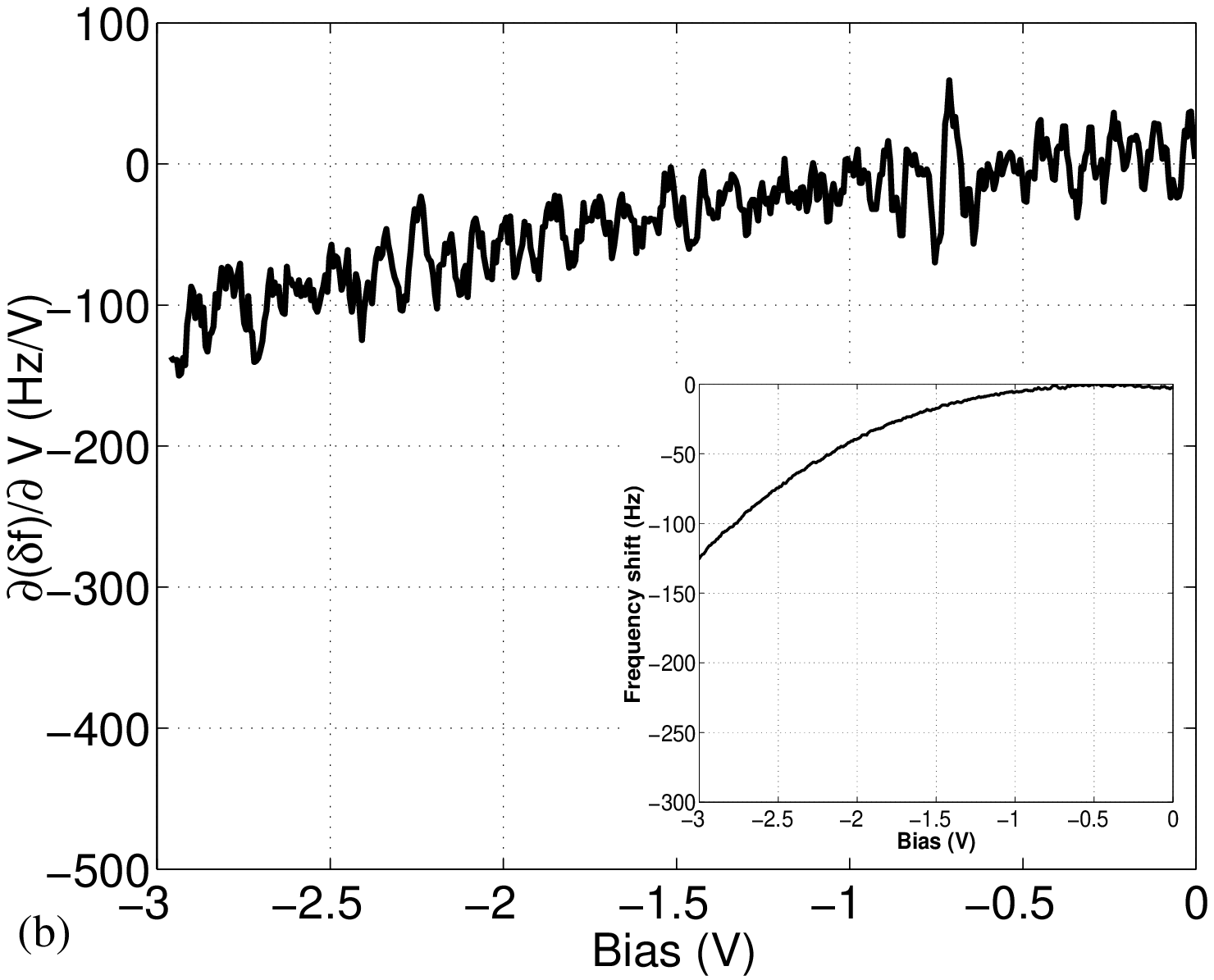}

\end{center}
\caption{Electrostatic force spectroscopy at (a) location A and
(b) location B of Fig. 1a . The numerical derivative of frequency
shift with respect to bias voltage, $\rm{{d(\delta f) }/{dV}}$ is
plotted versus bias voltage to clarify jumps in the cantilever
resonance frequency. Insets of (a) and (b) show the actual
frequency shift data on the quantum dot and on the wetting layer.
}\label{fig2}
\end{figure}

 The contact mode image shown in Fig. 1a shows two
locations where the EFS was performed at a tip-sample surface
separation of 13 nm. The obtained frequency shift data are plotted
in Fig. 2. Calibration of the tilt of the sample prior to EFS
measurements ensured that the tip-sample separation were identical
for these two points. The drift of the scanner was also measured
to be insignificant during the data acquisition period. Discrete
jumps in the cantilever frequency were observed on top of a QD
having a 40 nm base diameter and a height of 17 nm. These jumps
disappeared as the tip moved away from the QD in the plane of the
surface, while the height was kept constant.

\begin{figure}
\begin{center}
\epsfxsize=3in
\epsffile{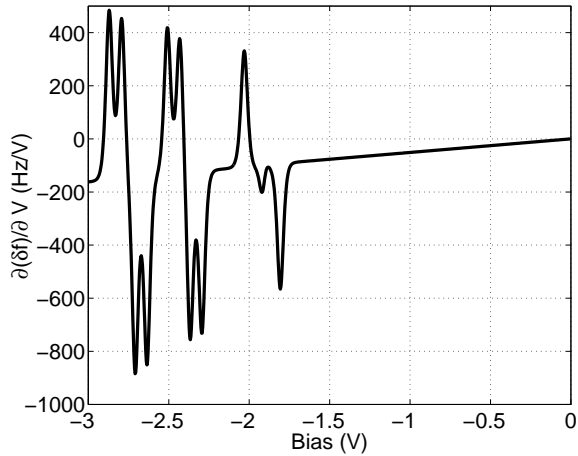}
\end{center}
\caption{The derivative of frequency shift with respect to bias
voltage, $\rm{{d(\delta f) }/{dV}}$ calculated from the theory
described in the text. Tip-sample separation and oscillation
amplitude used in calculation are 14.2 nm and 1 nm respectively.
Tip radius is assumed to be 20 nm. Electron energy levels are
obtained from the data presented in Table I . }\label{fig3}
\end{figure}

Based on the calculations of QD electron energy levels,
we expect to see two s-like states whose energy
degeneracy is lifted by an on-site Coulomb charging energy. In
addition, four p-like states whose energy degeneracies are lifted
both by on-site Coulomb charging effect and strain-induced anisotropy are
expected. According to calculations done for various sized QDs \cite{leburton}
, the on-site Coulomb charging energies for
s-like and p-like states differ and are estimated to be about 18
meV and 10 meV respectively, for 20 nm base diameter QDs in a
semi-insulating GaAs matrix.

\begin{table}
\caption{Electron energy levels inferred from previous
capacitance-voltage (CV) spectroscopy  for ensemble of 20 nm
average base diameter capped QDs and EFS experimental data
presented in this paper involving 40 nm base diameter uncapped QD.
Electron energies are in units of meV and are shifted to match the
ground state energies, $\rm{E_{s1}}$.}

\begin{tabular}{ccccc}
 &\multicolumn{2}{c}{Capacitance Measurement\tablenote{Ref. 4}}&\multicolumn{2}{c}{EFS}\\
 State&quantized &charging&quantized &charging \\

$\rm{E_{s1}} $&0&19 &0&9 \\
$\rm{E_{s2}} $&19& &9&  \\
$\rm{E_{px1}}$&74&8&40&6\\
$\rm{E_{px2}}$&82& &46& \\
$\rm{E_{py1}}$&100&10&68&6\\
$\rm{E_{py2}}$&110& &74& \\
\end{tabular}
\label{table}
\end{table}

The charging energies decrease to 11 meV and 5 meV for the s-like
and p-like states in 40 nm base diameter QDs. A first-principles
calculation \cite{jeongnimkim} that takes into account the strain
anisotropy in the QDs predicts the four p-like states to form two
pairs of states that are separated by about 25 meV.

Applying the above mentioned model to the experimental data shown
in Fig. 2, allows estimation of electron energy levels for the
first two states (s-like states) of 148 and 157 meV with respect
to the InAs conduction band minimum. The four p-like states are
also identified, and calculated to be 188, 194 , 216, and 222 meV
above the InAs conduction band minimum. Since band bending effects
have not been considered in the calculation, and the work function
difference of the tip and sample is known only approximately,
these energies are correct only to within an overall offset. Using
these QD energies in the electrostatic model described above, a
theoretical calculation of the expected frequency shift versus
bias voltage is plotted in Fig. 3. The data in Fig. 2a were
measured on top of a QD having a 40 nm base diameter. The observed
charging energies of 9 meV for the s-like states and 6 mev for the
p-like states in EFS data agrees well with theory in Ref. 10. A
comparison of electron energy levels for 20 nm base diameter InAs
QDs measured through conventional capacitance spectroscopy on
ensembles of QDs (Ref. 4) with results of EFS measurement on a
single 40 nm base QD is given in Table \ref{table}.

\section{Conclusion}

In conclusion, we have demonstrated an electrostatic force microscopy
technique for identifying the charge state of a single QD. Our
results on uncapped dots of 40 nm base diameter agree well with the theoretical
prediction for the charging energies for the s-like and p-like states and
and agree qualitatively with theoretical estimates of quantized energy levels
for this size QDs.
The EFS method presented has
the advantage that it can be applied without ultra-high vacuum (UHV)
conditions since the states probed can be below a capping layer.
It can also be extended to the study of states at interfaces of properly
designed MBE structures and of surfaces with adsorbates.

\section{Acknowledgments}

The authors would like to thank Glenn Solomon for valuable
discussions on InAs QDs.

\newpage

\end{document}